\documentclass[a4paper,11pt]{article}
\usepackage{pos}
\usepackage{booktabs}
\usepackage{multirow}
\usepackage{subcaption}

\title{Reducing Simulation Dependence in Neutrino Telescopes with Masked Point Transformers}

\author*[a]{Felix J. Yu}
\author[a]{Nicholas Kamp}
\author[a]{Carlos A. Arg\"{u}elles}

\affiliation[a]{Department of Physics \& Laboratory of Particle Physics and Cosmology, Harvard University,\\
Cambridge, MA, USA}

\emailAdd{felixyu@g.harvard.edu}
\emailAdd{nkamp@g.harvard.edu}
\emailAdd{carguelles@g.harvard.edu}

\abstract{Machine learning techniques in neutrino physics have traditionally relied on simulated data, which provides access to ground-truth labels. However, the accuracy of these simulations and the discrepancies between simulated and real data remain significant concerns, particularly for large-scale neutrino telescopes that operate in complex natural media. In recent years, self-supervised learning has emerged as a powerful paradigm for reducing dependence on labeled datasets. Here, we present the first self-supervised training pipeline for neutrino telescopes, leveraging point cloud transformers and masked autoencoders. By shifting the majority of training to real data, this approach minimizes reliance on simulations, thereby mitigating associated systematic uncertainties. This represents a fundamental departure from previous machine learning applications in neutrino telescopes, paving the way for substantial improvements in event reconstruction and classification.}

\ConferenceLogo{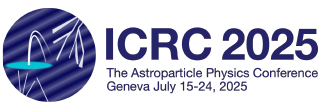}

\FullConference{39th International Cosmic Ray Conference (ICRC2025)\\
 15–24 July 2025\\
Geneva, Switzerland\\}

\begin{document}
\maketitle

\section{Introduction}\label{sec:intro}

Neutrino telescopes, such as IceCube, KM3NeT, and Baikal-GVD, have significantly advanced our understanding of high-energy astrophysical neutrinos. In recent times, machine learning (ML) models trained on detailed Monte Carlo (simulation) simulations have become increasingly prevalent. These models enable fast and accurate event reconstructions and classification. However, discrepancies between simulation and data that arise from complex environmental conditions, detector-specific systematics, and unmodeled effects, remain a persistent challenge. These discrepancies can yield biases in the reconstructions, or result in incorrect coverage assessments which can impact analysis conclusions.

Self-supervised learning (SSL) has emerged as a powerful framework for training ML models without labeled data. By leveraging the internal structure of unlabeled datasets, SSL enables models to learn useful representations through various tasks such as predicting masked inputs or distinguishing between data augmentations. This approach has gained traction in computer vision and natural language processing, where labeled datasets are costly or statistically limited. More recently, SSL has been explored in the context of neutrino telescopes~\cite{timiryasov2024polarbert} and proposed as a tool for domain adaptation~\cite{achituve2021self}.

Motivated by these advances, we present the first study to apply SSL for mitigating simulation dependence and mis-modeling effects in neutrino telescope reconstruction techniques, where labeled, simulated data are available but imperfect due to systematic uncertainties or unmodeled phenomena. The key idea is that SSL allows the majority of model training to be shifted onto unlabeled real data, thus bypassing the discrepancies. We focus on two representative tasks as benchmarks: directional reconstruction of muon neutrinos and separation of double cascades from tau neutrinos from single cascade backgrounds. Using a transformer architecture tailored for neutrino events, we compare conventional supervised models with SSL-based approaches. To evaluate robustness under realistic conditions, we introduce controlled mismatches between separated datasets by injecting uncorrelated noise hits. While this particular data-simulation discrepancy does not encompass the full spectrum of potential detector or physics mis-modeling effects, it serves as a representative and broadly relevant example of how simulation inaccuracies can affect performance. 

Our results show that supervised models remain robust when the noise is modeled but the rate is slightly incorrect, indicating resilience to moderate systematic errors. However, when the effect is entirely unmodeled, supervised models suffer significant performance degradation. In contrast, SSL maintains strong performance in both cases. This is particularly important in practice, where real detector data may include unknown phenomena that are not present in simulation. Therefore, our SSL method provides a valuable safeguard against such unmodeled discrepancies. The code for the models developed and used for this study are made available on \href{https://github.com/felixyu7/neptune/}{GitHub}.


\section{Method}\label{sec:arch}

\begin{figure}
    \centering
    \includegraphics[width=0.95\linewidth]{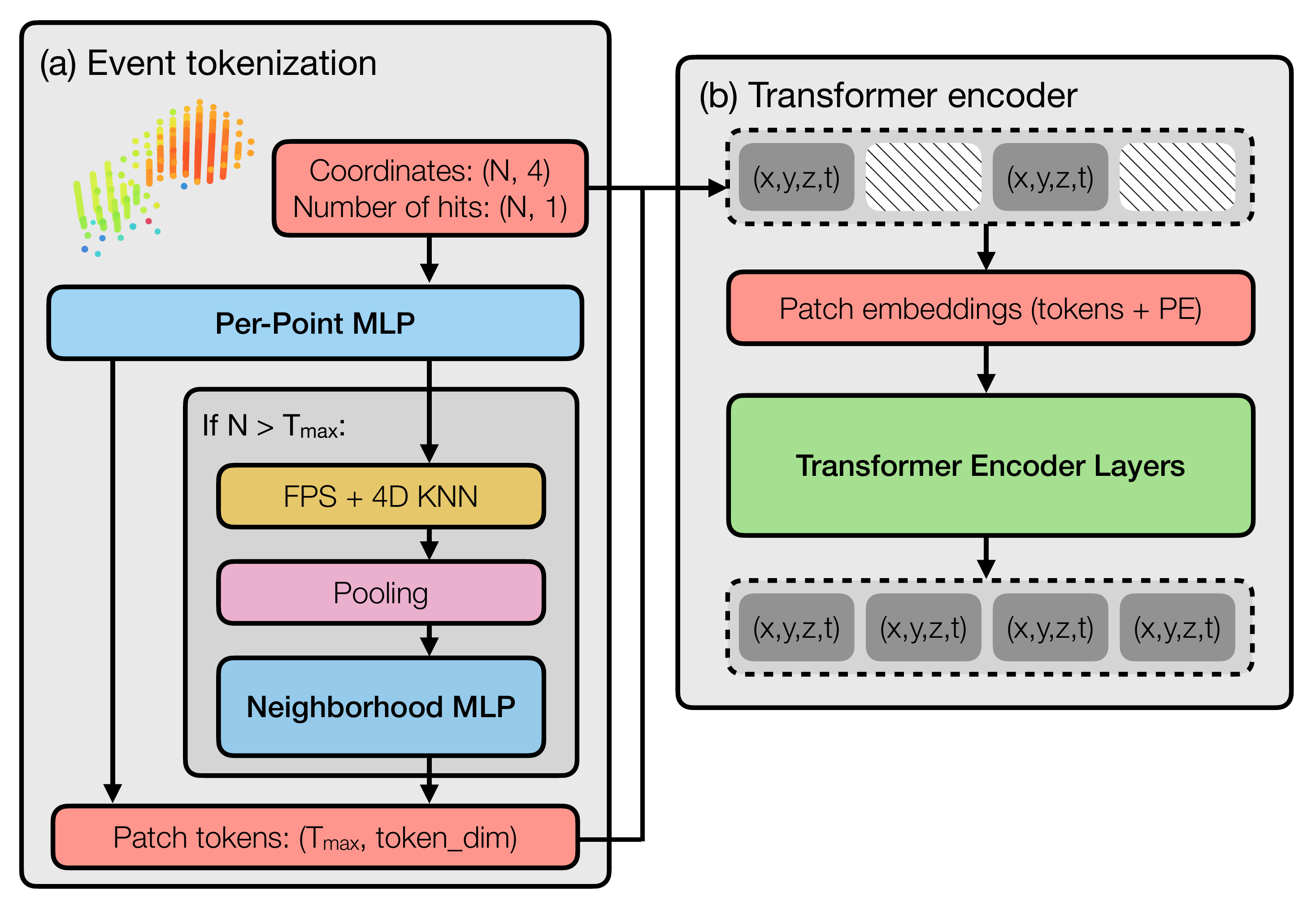}
    \caption{Architecture of the \texttt{neptune} model with masking.}
    \label{fig:arch}
\end{figure}

To serve as our backbone ML model for this study, we developed a transformer architecture grounded in point cloud methodologies, which we term a\textbf{N} \textbf{E}fficient \textbf{P}oint \textbf{T}ransformer for \textbf{U}ltrarelastivistic \textbf{N}eutrino \textbf{E}vents (\texttt{neptune}). The model consists of three main components: an event tokenizer, a transformer encoder, and a downstream task head. A high-level overview of the architecture is shown in Figure \ref{fig:arch}.

\subsection{Event Tokenizer}

Tokenization is the crucial initial step for transformer-based processing of neutrino telescope events, converting irregular raw data into token sequences with corresponding feature vectors suitable for transformers. Inspired by point cloud processing methods, our tokenizer achieves efficient inference even on CPUs, a key advantage given the vast data volumes and high event rates of gigaton-scale neutrino telescopes.

The raw input data consists of four-dimensional spatio-temporal coordinates per sensor hit, along with associated feature vectors. Hits are aggregated into 3~ns time windows, reflecting typical digitizer sampling resolutions in photomultiplier tubes. Each feature encodes the number of hits per bin. This approach can also easily incorporate the \texttt{om2vec} method \cite{Yu:2024eog}, which employs a variational autoencoder to encode OM-level timing information, allowing for a more efficient three-dimensional event representation with minimal impact on performance.

As illustrated in Figure~\ref{fig:arch}, our tokenizer follows a PointNet-inspired strategy \cite{qi2017pointnet, zhao2021point}, initially applying a per-point multilayer perceptron (MLP) to each optical module's (OM) features. Subsequent steps depend on the point cloud size relative to a predefined maximum sequence length, $T_{\text{max}}$, for transformer processing. If the number of hit sensors, $N$, exceeds $T_{\text{max}}$, farthest point sampling (FPS) reduces the size to maximize spatial coverage. Spatial and temporal context is then incorporated via four-dimensional k-nearest neighbors (KNN) searches, extracting and pooling neighbor features through an additional MLP. For events with fewer hits, FPS and KNN are omitted, and sequences are padded. In both cases, the output is a sequence of patch tokens with length $T_{\text{max}}$. This design enables \texttt{neptune} to efficiently handle variable event sizes. Given the large number of photon hits in high-energy events, we set $T_{\text{max}}=512$ and $k=32$.

\subsection{Encoder}

The encoder primarily consists of transformer blocks, processing sequences derived from tokenizer-generated patch token features. Tokens are enriched with spatial positions and first-hit times of each OM (or centroid space-time positions if FPS is applied), providing essential context. For supervised learning and fine-tuning scenarios, transformer outputs are aggregated via mean pooling and passed to a downstream prediction head.

\subsection{Pre-training}

Leveraging recent advances in masked image modeling and masked autoencoders \cite{he2022masked, pang2022masked, Young:2025qah}, our pre-training approach involves reconstructing masked input components. Masking occurs post-tokenization and pre-transformer encoding, and is illustrated in Figure~\ref{fig:arch}.

We randomly mask spatio-temporal coordinates and train the transformer to reconstruct them using smooth L1 loss. Several masking schemes were evaluated, including spatial-only, temporal-only, and combined spatio-temporal masking. Though optimal strategies slightly varied depending on the downstream task, differences were minor. Aggressive masking ratios between 0.75 and 1.0 were applied based on retained information type. These high masking ratios create challenging reconstruction tasks, encouraging the model to learn the inherent structure of neutrino data without explicit labels. 

\subsection{Finetuning}

Following pre-training, a prediction head is attached, and the model undergoes finetuning with labeled data for reconstruction tasks. We adopted the block expansion \cite{bafghi2024parameter} finetuning method for this work, which inserts identity-initialized transformer blocks atop frozen pre-trained layers. These blocks, exclusively trained on downstream tasks, preserve initial outputs, mitigating catastrophic forgetting of the target domain during finetuning. This is particularly beneficial in our context, where robustness to domain shifts between data and simulation is critical.

\section{Experiments}\label{sec:experiments}

In this section, we setup a case study using two simulated datasets: one treated as ``data'' and the other as labeled Monte Carlo (simulation). To mimic realistic experimental conditions, we use only the labels from the simulation set for training and introduce controlled discrepancies between the datasets by injecting uncorrelated noise hits.

\subsection{Noise Hit Generation}

\begin{figure}
    \centering
    \includegraphics[width=0.75\linewidth]{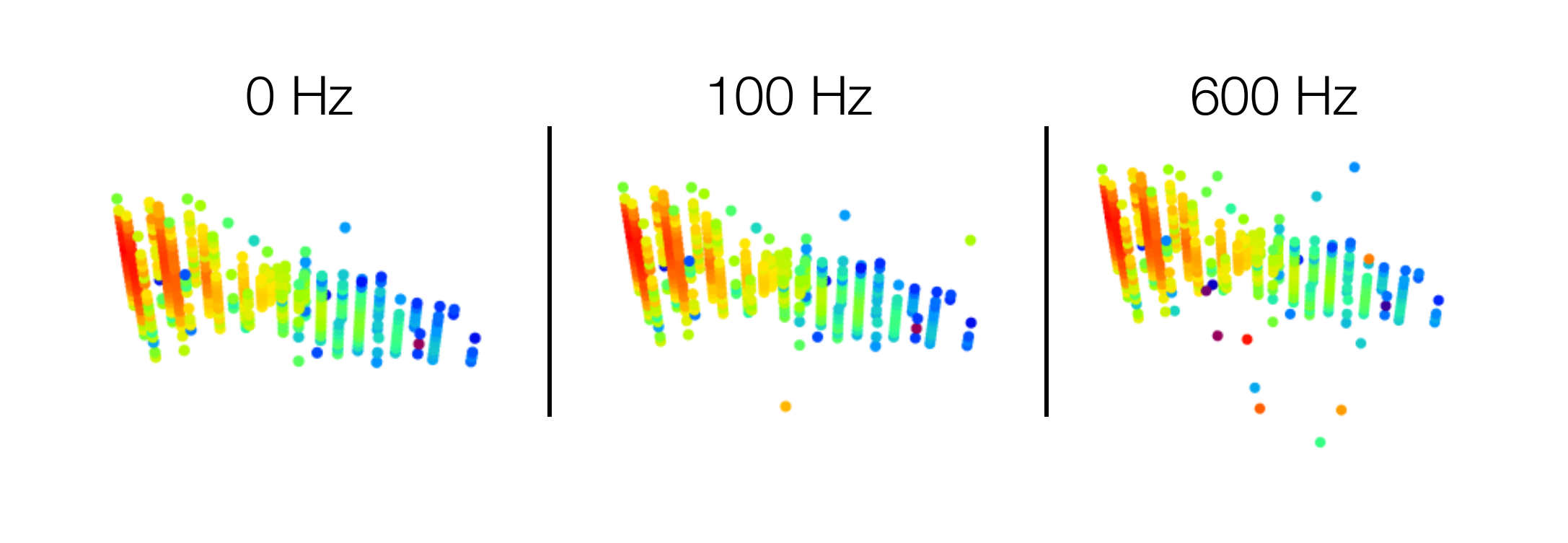}
    \caption{Example track-like event with varying levels of uncorrelated noise hits. Each point represents an OM that registered a photon hit, with color indicating the first-hit time (red = early, blue = late).}
    \label{fig:noise}
\end{figure}

To simulate uncorrelated noise, we inject random hits into each optical module (OM) at a fixed rate \( R \) (e.g., 500 Hz)~\cite{larson2013simulation}. For each OM \( i = 1, \dots, N \), located at position \( \mathbf{r}_i = (x_i, y_i, z_i) \), we sample the number of noise hits \( n_i \) from a Poisson distribution with mean \( \lambda = (R / 10^9) \cdot \Delta t \), where \( \Delta t = t_1 - t_0 \) is the duration of the injection window in nanoseconds. Each of the \( n_i \) hits is placed at position \( \mathbf{r}_i \) with a timestamp drawn uniformly from the interval \( [t_0, t_1] \). The full set of noise hits is then concatenated with the original event hits. In our setup, \( t_0 \) is the event start time and \( t_1 \) is the event end time plus 1000~ns. Figure~\ref{fig:noise} illustrates an example event at different noise levels.

\subsection{Datasets \& Models}

All datasets in our experiments are generated using the open-source neutrino telescope simulation framework \texttt{Prometheus} \cite{Lazar:2023rol}. We adopt an IceCube-like configuration with an ice medium and a similarly arranged OM geometry, though the proposed methods are readily transferable to alternative detector designs, including those based in water. We examine two downstream tasks that are central to neutrino telescope analyses: (1) directional reconstruction, which is critical for most physics analyses, and (2) single versus double cascade classification, which is important for $\nu_{\tau}$ identification. 

For directional reconstruction, we simulate track-like events ($\nu_\mu$ CC), and divide them into three subsets: 750{,}000 events for ``data'', 250{,}000 for simulation, and an additional 120{,}000 events for validation and testing. For the cascade classification task, we simulate a total of 200{,}000 single cascade events (from $\nu_e$ CC and $\nu_\mu$ NC) and approximately 10{,}000 double cascade events (from $\nu_\tau$ CC with a tau lepton propagation distance greater than 10 meters). All datasets are generated with an $E^{-1}$ energy spectrum, ranging from 100~GeV to 1~PeV.

For each task, we train two models: an SSL model that is pre-trained on ``data'' and fine-tuned on labeled simulation, and a baseline supervised model trained directly on labeled simulation. The SSL pre-training objective is task-specific: for directional reconstruction we use \emph{Temporal Masking} (ratio 1.0), and for cascade classification we use \emph{Spatio-temporal Masking} (ratio 0.75). In both cases, we study robustness to uncorrelated noise hits by considering two evaluation scenarios:

\begin{itemize}
    \item \textbf{Un-modeled noise:} simulation contains no uncorrelated noise hits (0~Hz), while the ``data'' contain noise. The per-task ``data'' noise rates are 100~Hz for directional reconstruction and 600~Hz for cascade classification.
    \item \textbf{Varying noise rates:} both domains contain noise with a modest mismatch: 600~Hz in ``data'' and 500~Hz in simulation (for both tasks).
\end{itemize}

All experiments use the same \texttt{neptune} architecture and hyperparameters. To ensure a fair comparison, supervised and SSL models were given equal exposure to labeled simulation data by matching supervised training time to SSL fine-tuning time. For directional reconstruction, the supervised model was trained for 30 epochs, while the SSL model was pre-trained on ``data'' for 30 epochs and fine-tuned on simulation for 30 epochs. For cascade classification, SSL used 20 epochs of pre-training and 10 epochs of fine-tuning, with the supervised model trained for 10 epochs on labeled simulation.

\section{Results}\label{sec:results}

\subsection{Un-modeled noise}

\begin{table}[h]
\centering
\captionsetup[subtable]{justification=centering}
\begingroup
\small
\setlength{\tabcolsep}{5pt}
\begin{subtable}[t]{0.48\textwidth}
\centering
\caption{Directional Reconstruction}
\label{tab:unmodeled_noise_dir}
\begin{tabular}{lcc}
\midrule
\textbf{Model} & \textbf{Med. $\Delta \phi$ (``data'')} & \textbf{Med. $\Delta \phi$ (sim)} \\
\midrule
Supervised & 20.50 & 2.153 \\
SSL        & \textbf{4.999} & \textbf{2.098} \\
\midrule
\end{tabular}
\end{subtable}
\hfill
\begin{subtable}[t]{0.48\textwidth}
\centering
\caption{Cascade Classification}
\label{tab:unmodeled_noise_casc}
\begin{tabular}{lcc}
\midrule
\textbf{Model} & \textbf{PR-AUC (``data'')} & \textbf{PR-AUC (sim)} \\
\midrule
Supervised & 0.226 & \textbf{0.364} \\
SSL        & \textbf{0.287} & 0.357 \\
\midrule
\end{tabular}
\end{subtable}
\caption{Un-modeled noise scenario. Simulation has 0~Hz noise; ``data'' has 100~Hz (directional reconstruction) and 600~Hz (cascade classification).}
\label{tab:unmodeled_noise}
\endgroup
\end{table}

\begin{figure}[t]
    \centering
    \includegraphics[width=0.8\linewidth]{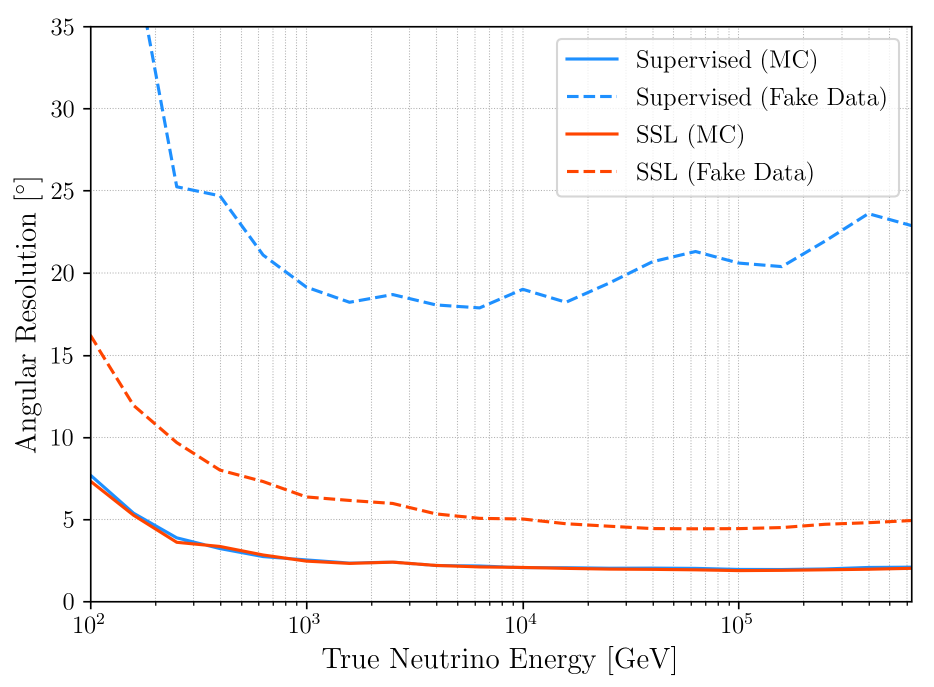}
    \caption{Directional reconstruction results as a function of neutrino energy in the un-modeled noise scenario.}
    \label{fig:angular_resolution}
\end{figure}

Table~\ref{tab:unmodeled_noise} shows that when simulation lacks noise hits entirely, the supervised model collapses on ``data'' in both the directional reconstruction and cascade classification tasks. SSL, pre-trained on noisy ``data'', remains robust: for directional reconstruction both models achieve $\sim 2^{\circ}$ on simulation, but on ``data'' the SSL model degrades only to $5^{\circ}$ while the supervised model drops to $20.5^{\circ}$. For cascade classification, the supervised model suffers a marked performance drop on ``data'', whereas SSL generalizes better across the domain shift. Figure~\ref{fig:angular_resolution} shows the directional reconstruction results as a function of energy, highlighting consistent performance degradation across the full range and a pronounced gap between the SSL and supervised models when applied to ``data''.

\subsection{Varying noise rates}

\begin{table}[h]
\centering
\captionsetup[subtable]{justification=centering}
\begingroup
\small
\setlength{\tabcolsep}{5pt}
\begin{subtable}[t]{0.48\textwidth}
\centering
\caption{Directional Reconstruction}
\label{tab:varying_noise_dir}
\begin{tabular}{lcc}
\midrule
\textbf{Model} & \textbf{Med. $\Delta \phi$ (``data'')} & \textbf{Med. $\Delta \phi$ (sim)} \\
\midrule
Supervised & \textbf{2.269} & \textbf{2.248} \\
SSL        & 2.343 & 2.321 \\
\midrule
\end{tabular}
\end{subtable}
\hfill
\begin{subtable}[t]{0.48\textwidth}
\centering
\caption{Cascade Classification}
\label{tab:varying_noise_casc}
\begin{tabular}{lcc}
\midrule
\textbf{Model} & \textbf{PR-AUC (``data'')} & \textbf{PR-AUC (sim)} \\
\midrule
Supervised & \textbf{0.338} & \textbf{0.348} \\
SSL        & 0.333 & 0.341 \\
\midrule
\end{tabular}
\end{subtable}
\caption{Varying noise rates scenario. Both domains contain noise with a modest mismatch: 600~Hz in ``data'' and 500~Hz in simulation (both tasks).}
\label{tab:varying_noise}
\endgroup
\end{table}

With 600~Hz noise in ``data'' and 500~Hz in simulation (Table~\ref{tab:varying_noise}), the supervised model is robust to the modest mismatch and performs comparably to SSL for directional reconstruction. For cascade classification, both approaches perform similarly on both domains, indicating that moderate systematic mis-modeling in noise rates is tolerable and does not lead to catastrophic degradation.

\section{Conclusion}\label{sec:conclusion}

In this proceeding, we have explored the benefits of integrating self-supervised learning (SSL) techniques into neutrino telescopes to reduce dependence on simulations. Our systematic experiments reveal that while traditional supervised ML methods can handle small, known systematic discrepancies adequately, they exhibit significant performance degradation when encountering subtle, un-modeled effects absent in training simulations. In contrast, SSL-based methods, which leverage realistic but unlabeled data, demonstrate substantial robustness and maintain stable performance even in the presence of unknown systematics.

Our findings underscore the importance and effectiveness of SSL in safeguarding neutrino telescope reconstruction techniques against unforeseen discrepancies between simulations and real data. Future work will explore the deployment of this approach on real experimental data, with a particular focus on assessing robustness in large-scale detectors such as IceCube.

\section{Acknowledgments}
This work is supported by the National Science Foundation under Cooperative Agreement PHY-2019786 (The NSF AI Institute for Artificial Intelligence and Fundamental Interactions, http://iaifi.org/.

\setlength{\bibsep}{0pt plus 0.3ex}


\begin{thebibliography}{99}

\bibitem{timiryasov2024polarbert}
I.~Timiryasov, J.-L.~Tastet, and O.~Ruchayskiy,
``PolarBERT: A Foundation Model for IceCube,''
in \textit{Proceedings of the Machine Learning and the Physical Sciences Workshop at NeurIPS 2024}, 2024.

\bibitem{achituve2021self}
I.~Achituve, H.~Maron, and G.~Chechik,
``Self-Supervised Learning for Domain Adaptation on Point Clouds,''
in \textit{Proceedings of the IEEE/CVF Winter Conference on Applications of Computer Vision (WACV)},
2021, pp. 123--133.

\bibitem{Yu:2024eog}
F.~J.~Yu, N.~Kamp and C.~A.~Arg{\"u}elles,
[arXiv:2410.13148 [physics.data-an]].

\bibitem{qi2017pointnet}
C.~R. Qi, H.~Su, K.~Mo, and L.~J. Guibas,
``PointNet: Deep learning on point sets for 3D classification and segmentation,''
in \textit{Proceedings of the IEEE Conference on Computer Vision and Pattern Recognition (CVPR)}, 
2017, pp. 652--660.

\bibitem{zhao2021point}
H.~Zhao, L.~Jiang, J.~Jia, P.~H.~S. Torr, and V.~Koltun,
``Point Transformer,''
in \textit{Proceedings of the IEEE/CVF International Conference on Computer Vision (ICCV)},
2021, pp. 16259--16268.

\bibitem{he2022masked}
K.~He, X.~Chen, S.~Xie, Y.~Li, P.~Dollár, and R.~Girshick,
``Masked Autoencoders Are Scalable Vision Learners,''
in \textit{Proceedings of the IEEE/CVF Conference on Computer Vision and Pattern Recognition (CVPR)},
2022, pp. 15979--15988.

\bibitem{pang2022masked}
Y.~Pang, W.~Wang, F.~E.~H. Tay, W.~Liu, Y.~Tian, and L.~Yuan,
``Masked Autoencoders for Point Cloud Self-supervised Learning,''
in \textit{Computer Vision – ECCV 2022}, Lecture Notes in Computer Science, vol. 13662, Springer, Cham, 2022, pp. 604--621.

\bibitem{Young:2025qah}
S.~Young, Y.~j.~Jwa and K.~Terao,
[arXiv:2502.02558 [hep-ex]].

\bibitem{bafghi2024parameter}
R.~A. Bafghi, N.~Harilal, C.~Monteleoni, and M.~Raissi,
``Parameter Efficient Fine-tuning of Self-supervised ViTs without Catastrophic Forgetting,''
in \textit{Proceedings of the IEEE/CVF Conference on Computer Vision and Pattern Recognition (CVPR) Workshops},
June 2024, pp. 3679--3684.

\bibitem{larson2013simulation}
M.~J. Larson,
``Simulation and Identification of Non-Poissonian Noise in the IceCube Neutrino Observatory,''
Master's thesis, University of Alabama, December 2013.

\bibitem{Lazar:2023rol}
J.~Lazar, S.~Meighen-Berger, C.~Haack, D.~Kim, S.~Giner and C.~A.~Arg{\"u}elles,
Comput. Phys. Commun. \textbf{304}, 109298 (2024)
doi:10.1016/j.cpc.2024.109298
[arXiv:2304.14526 [hep-ex]].

\end{thebibliography}
\end{document}